\documentclass[prb,twocolumn,english,superscriptaddress]{revtex4-1}
\usepackage{amsmath,amssymb,mathrsfs}

\usepackage{amsfonts,bm}
\usepackage{amsmath}
\usepackage{amssymb}
\usepackage{amsthm}
\usepackage{graphicx}
\usepackage{graphics}
\usepackage{subfigure}
\usepackage{color}
\usepackage{bm}
\usepackage{hyperref}
\usepackage{rotating}
\usepackage{comment}

\newcommand{\be}{\begin{equation}}
\newcommand{\ee}{\end{equation}}
\newcommand{\bea}{\begin{eqnarray}}
\newcommand{\eea}{\end{eqnarray}}

\newcommand{\abs}[1]{\left| #1 \right|} % for absolute value
\newcommand{\ket}[1]{\left| #1 \right>} % for Dirac bras
\newcommand{\bra}[1]{\left< #1 \right|} % for Dirac kets
\newcommand{\braket}[2]{\left< #1 \vphantom{#2} \right| \left. #2 \vphantom{#1} \right>} % for Dirac brackets
\newcommand{\matrixel}[3]{\left< #1 \vphantom{#2#3} \right| #2 \left| #3 \vphantom{#1#2} \right>} % for Dirac matrix elements
\let\baraccent=\= % rename builtin command \= to \baraccent

\newcommand{\ba}{\begin{eqnarray} }
\newcommand{\ea}{\end{eqnarray} }

\newcommand{\nn}{\nonumber}

\newcommand{\bpm}{\begin{pmatrix}}
\newcommand{\epm}{\end{pmatrix}}
\newcommand{\bmm}{\begin{matrix}}
\newcommand{\emm}{\end{matrix}}

\makeatother

\usepackage{babel}
\begin{document}

\title{Floquet dynamics of disordered bands with isolated critical energies}
 \author{Sriram Ganeshan}
 \affiliation{Department of Physics, City College, City University of New York, New York, NY 10031, USA }
 \author{Kartiek Agarwal}
\affiliation{Department of Electrical Engineering, Princeton University, Princeton NJ 08544, USA}
\affiliation{Department of Physics, McGill University, Montr\'eal, Qu\'ebec H3A 2T8, Canada}
\author{R. N. Bhatt}
\affiliation{Department of Electrical Engineering, Princeton University, Princeton NJ 08544, USA}
\affiliation{School of Natural Sciences, Institute for Advanced Study, Princeton NJ 08540, USA}

\date{\today}

%%%%%%%%%%%%%
\begin{abstract}
We investigate localization properties of driven models which exhibit a sub-extensive number of extended states in the static setting. We consider instances where the extended modes are or are not protected by topological considerations. To this end, we contrast the strongly driven disordered lowest Landau level, which we refer to as the random Landau model (RLM), with the random dimer model (RDM); the latter also has a sub-extensive set of delocalized modes in the middle of the spectrum whose origin is not topological.  We map the driven models on to a higher dimensional effective model and numerically compute the localization length as a function of disorder strength, drive amplitude and frequency using the recursive Green's function method. Our numerical results indicate that in the presence of a strong drive (low frequency and/or large drive amplitude), the topologically protected RLM continues to exhibit a spectrum with both localized and delocalized (or critical) modes, but the spectral range of delocalized modes is enhanced by the driving. This occurs due to an admixture of the localized modes with extended modes arising due to the topologically protected critical energy in the middle of the spectrum. On the other hand, in the RDM, a weak drive immediately localizes the entire spectrum. This occurs in contrast to the naive expectation from perturbation theory that mixing between localized and delocalized modes generically enhances the delocalization of all modes. Our work highlights the importance of the origin of the delocalized modes in the localization properties of the corresponding Floquet model.
\end{abstract}
%%%%%%%%%%%%%
\maketitle

%%%%%%%%%%%%%%%%%%%%%%%%
%%%%%%%%%%%%%%%%%%%%%%%%
\section{Introduction} 

Floquet dynamics~\cite{floquet1883equations, floquet1884equations} is a subclass of unitary dynamics wherein systems are subject to a periodic-in-time external drive. Recent work has shown that a new class of dynamical phases which have no static analogue can arise in the Floquet setting in quantum systems~\cite{kitagawa2011transport, lindner2013topological, khemani2016phase, yao2017discrete, titum2016anomalous, higashikawa2019floquet, peri2018anomalous}. A priori this is unexpected since driven systems are expected to generically heat up to infinite temperature~\cite{lazaridesheating}, which further implies, by virtue of the eigenstate thermalization hypothesis, that the Floquet unitary must itself be featureless~\cite{moessner2017equilibration}. However, in certain instances, novel quantum dynamics may arise when heating is prevented due to the presence of strong disorder and many-body localization~\cite{time_crystals_shivaji, khemani2016phase, von2016absolute,yao2017discrete} (MBL) or the presence of topologically protected modes which remain decoupled from the bulk in the Floquet setting~\cite{oka2009photovoltaic, inoue2010photoinduced, liangmajoranazeropi, kitagawa2011transport,  lindner2011floquet, lindner2013topological, rudner2013anomalous,  titum2016anomalous, roy2017floquet, roy2017periodic, potter2016classification}. Additionally, when the system is driven at high-enough frequencies, a prethermal regime may be realized wherein an effective Hamiltonian describes approximate unitary dynamics at stroboscopic (and related) times for an exponentially long timescale~\cite{abanin2015exponentially, bukov2015prethermal, bukov2016heating, abanin2017effective, time_crystals_dominic, von2016absolute, dumitrescu2018logarithmically}. %Time crystals are the prototypical example of such a phase which is an interplay of  spontaneous symmetry breaking, many body localization and Floquet driving. 
Some of these Floquet phases, for instance time crystals, have been recently realized across several experimental platforms~\cite{zhang2017observation,   lukin2017observation, barends2013coherent, rovny2018observation}. 
%Another direction of pursuit has been the study of Floquet phases that exhibit novel topological invariants that arise due to the periodicity in quasi-energies~\cite{oka2009photovoltaic, inoue2010photoinduced, kitagawa2011transport,  lindner2011floquet, lindner2013topological, rudner2013anomalous,  titum2016anomalous, roy2017floquet, roy2017periodic, potter2016classification}. 

Particularly in the context of low-frequency driving, where prethermalization is absent, an important direction of investigation is the study of the transition between systems that eschew heating and espouse novel quantum dynamics, to those in which resonances proliferate, leading to the delocalization and eventual heating of the system. In this regard, several works have considered the mechanism for the proliferation of resonances in systems with a fully localized set of states by virtue of driving, both in the interacting, Floquet-MBL setting~\cite{abanin2016theory,pontefloquetMBL,ponte2015periodically}, and the single-particle Anderson insulator setting~\cite{agarwal2017localization}. Less is known about if and how such resonances manifest in systems which have a mix of both localized and delocalized states, although a previous numerical study~\cite{lazarides2015fate} suggests that in interacting systems with a mobility edge, delocalization occurs immediately at infinitesimally weak driving. 
%
%% In the former, interactions lead to delocalization when driving at low-enough frequencies, while in the latter, in particular, despite the presence of resonances, driving does not generically lead to delocalization.   
%
%A natural question then arises if this scenario persists in a system that is not fully localized, via a mobility edge or a critically delocalized state.  Investigations into driving systems with mobility edges, as in Ref., reveals immediate delocalization and heating even for weak driving. 
%In the above settings, heating is expected to occur exponentially slowly, and an effective (and local) Hamiltonian description can thus be formulated. 
Such a result follows naturally from perturbative arguments: effective hybridization of localized modes with the delocalized mode should lead to delocalization in general. At the same time, localization itself follows from non-perturbative effects, and given its robustness in low dimensions, it is not obvious such perturbative arguments always apply straightforwardly, especially in the low-frequency, ``strong driving" regime. Thus, understanding how resonances lead to delocalization in systems of mixed localized and delocalized states requires detailed investigation.  

%This is particularly true in the strong-driving setting 

In this work, we take a step towards the above goal by numerically studying localization in Floquet systems which, in the static setting, exhibit a sub-extensive set of delocalized states, located around a critical energy at which the localization length diverges. Such models represent a controlled departure from the setting of fully localized models previously studied~\cite{agarwal2017localization}, and further have natural appeal as they arise in many physical systems, particularly in the quantum Hall setting. Furthermore, we make a distinction between models where the extended states arise due to topological considerations as opposed to those in which these states arise due to non-topological reasons. In order to access the physics of such extended states and study their properties in the driven setting, we focus on non-interacting models. 

 Concretely, we study two models with extended states arising with/without topological underpinning. For the topological case, we consider the Floquet dynamics of electrons in the lowest Landau level along with some level broadening due to local potential disorder, which we refer to as the random Landau model (RLM). The undriven case has been considered by several authors~\cite{aoki1985critical, huckestein1990one, huo1992current}“; Huo and Bhatt \cite{huo1992current}, in particular, have shown that the critical energy corresponds to a sub-extensive set of delocalized states. For the non-topological case, we consider random dimer model (RDM) which, like the RLM, has states with a diverging localization length near a single energy in the spectrum~\cite{dunlap1990absence} . We numerically compute the localization length as a function of the drive parameters and disorder strength. Our numerics indicate that in the case of the RLM, driving results in \emph{enhanced} delocalization of modes near the critical energy but there remain localized modes at the edges of the band for even large drive amplitudes. Our result may be understood as a consequence of the fact that the Chern number of the isolated band studied cannot be changed by the driving, which guarantees the presence of an extended state by Laughlin's argument~\cite{laughlingquantized}. In contrast, for the driven RDM, the delocalized states become localized even in the presence of a weak drive. We thus provide evidence that delocalization due to periodic driving in the presence of few delocalized states is not a given and is subject to the origin of these delocalized modes. For the latter, a simple estimate for the mean-free path obtained due to scattering by time-dependent variations of the drive which modifies the local potential, agrees reasonably well with the Floquet localization length. 

%Most of the works highlighted above have either focussed on the high frequency regime of the periodic drive or assumed there is some mechanism opening up a long pre-thermal time scale. Furthermore, it has been assumed that there is no mobility edge in the spectrum. It has been numerically shown that driving the quantum random energy model with a mobility edge the system delocalizes even with a weak drive~\cite{lazarides2015fate}. However, there is no reason for the delocalization to be the generic scenario and could very well be a model dependent effect.

%The strong driving regime or the low frequency regime in particular manifests resonances which can proliferate in the system, thereby forbidding a perturbative approach. 
The numerical study of the low frequency, or strong driving regime is made tractable by mapping our d-dimensional time dependent Hamiltonian onto a $(d+n)$-dimensional system with $n$ harmonic space directions in a manner similar to Ref.~\onlinecite{shirley1965solution} (for single frequency driving considered in this work, the harmonic space dimension $n=1$). The localization properties of this effective Hamiltonian can then be studied using the recursive Green's function approach.  For the effective Floquet Hamiltonian in $(d+1)$ dimensional space, the authors showed in Ref.~\onlinecite{agarwal2017localization} that the method gives the effective localization length of the Floquet eigenstates.   %Recent works have shown that real space topological insulators can be realized in fully temporal lattices~\cite{martin2017topological}. This method was applied to the case of Bernevig-Hughes-Zhang (BHZ) model in 2D where the topological lattice was created in the resonance space by driving a spin with two periodic but mutually incommensurate frequencies. 
%The authors of this paper have recently showed that strong Floquet driving of the one dimensional Anderson insulator does not result in heating~\cite{agarwal2017localization}. 
%Thus the strong periodic driving of disordered systems may open new directions in the search for robust dynamical phases.  

This paper is organized as follows. We begin with a review of Floquet lattice mapping of a periodically driven system and outline the recursive Green's function method to calculate localization length for the Floquet lattice in the presence of static disorder. We then apply this method to two models with a single delocalized state in the thermodynamic limit, namely, Random Dimer Model (RDM) and Random Landau Model (RLM) and discuss our numerical findings. We conclude by discussing the qualitative differences between these two cases and outline future directions.

\section{Floquet Hamiltonian in higher dimensional space and transfer matrix method}
Quantum Hamiltonians subject to a time dependent periodic drive of arbitrary strength can be treated using the methods of Ref.~\onlinecite{shirley1965solution} which studies the monochromatic driving of a spin-$1/2$ system by mapping it to one higher dimension. We note in passing that a single spin driven by multiple, incommensurate frequencies was treated in an analogous fashion by mapping it to a system in as many dimensions as the number of driving frequencies in Ref.~\onlinecite{martin2017topological}. For the purposes of this work, we are interested in driving a $d$ dimensional system at a single frequency. This can be studied by mapping the system using the above methods to a $(d+1)$-dimensional Hamiltonian in an effective electric field. We will next review this mapping and the recursive Green's function method used to compute localization properties in static models, before finally discussing how the method can be adapted to the Floquet setting to determine Floquet localization lengths. 
%The lattice coordinate of the extra dimension keeps track of the number of energy quanta absorbed from the drive; alternatively, one may view this dimension as corresponding to copies of the static Hilbert space, for different photon number $n$. 

\subsection{Review of the Floquet lattice construction}

We now briefly review the Floquet lattice mapping of a periodically-driven Hamiltonian. Consider first a generic periodic-in-time Hamiltonian $H (\omega t) $, where $\omega$ is the driving frequency, written in the basis of states $\ket{i}$ which correspond to the \emph{physical} Hilbert space of the static part of the Hamiltonian---
\begin{align}
	H=\sum_{jk}h^{jk}(\omega t)|j\rangle\langle k|,
\end{align}
where $h^{jk}(\omega t)=\langle j |H_0+V(\omega t)|k\rangle$, with $H_0$ being the static term and $V(\omega t)$ is the time dependent driving term. Floquet eigenstates of this system can be found by expanding the time-dependent wavefunction as a Fourier series in the temporal harmonics: 
\begin{align}
\ket{\psi (t)}=e^{-i\epsilon t}\sum_{j,n}\phi_{j,n} e^{-i n\omega t} \ket{j},
\label{eq:floqeig}
\end{align}
and whose coefficients satisfy the equation 
\begin{align}
	(\epsilon_{\alpha}+n \omega \phi_{j,n})=\sum_{m}  h^{jk}_m  \phi_{k,n-m}, \nonumber \\
	h^{jk} (\omega t) =\sum_{m} h^{jk}_m e^{-i m\omega t}. 
	\label{eq:qe}
\end{align}

In Eq.~(\ref{eq:qe}), $\epsilon$ is the quasi-energy corresponding to the Floquet eigenstate $\ket{\psi}$. The above equation can in fact be thought of as an effective Schr\"odinger equation operating in a $(d+1)$ dimensional lattice system with the temporal harmonic number being an extra, discrete dimension. $h^{ij}_m$ serve as effective hopping parameters which not only describe hopping from states $\ket{i, n} \rightarrow \ket{j, n+ m}, \forall \ket{n} $, reflecting the absorption of $m$ photons by the system while hopping from physical basis state $\ket{i}$ to state $\ket{j}$. The extra photons cost an energy $n \omega$ as one would naively expect; this can be interpreted as an effective electric field operating in the harmonic-space direction. 

Note also that the above equations have a redundancy: there are only $L$ unique Floquet eigenstates, where $L$ is the the Hilbert space dimension of the physical system. The other Floquet eigenstates have a quasi-energy that is related to these unique states with a shift of a multiple of the drive frequency. Thus, we may denote different Floquet eigenstates $\ket{\psi_{\alpha, n} (t)}$ by the indexes $\alpha$, and $n$; these have quasi-energies $\epsilon = \epsilon_\alpha + n \omega$.  

\subsection{\label{sec:rgf} Review of the recursive Green's function approach for computing the localization length}
We would like to study the localization length of Floquet eigenstates. This will be done using a straightforward generalization of the recursive Green's function approach for computing the localization length in static systems, to which we now turn. 

%Studying localization or delocalization on the Floquet lattice is closely related to studying whether the system in question can continuously absorb energy from the drive and heat to infinite temperatures. This question is non-trivial if we consider the static term in the Hamiltonian $H_0$ to be disordered. The interplay of static disorder and drive can result in the system being localized on the Floquet lattice which in turn implies energy localization. In order to study the localization properties on the Floquet lattice, we employ the transfer matrix  or 
%%%%%%%%%%%%%%%%%%%%%%%%%%%%%%
\begin{center}
\begin{figure}
\includegraphics[width=0.4\textwidth]{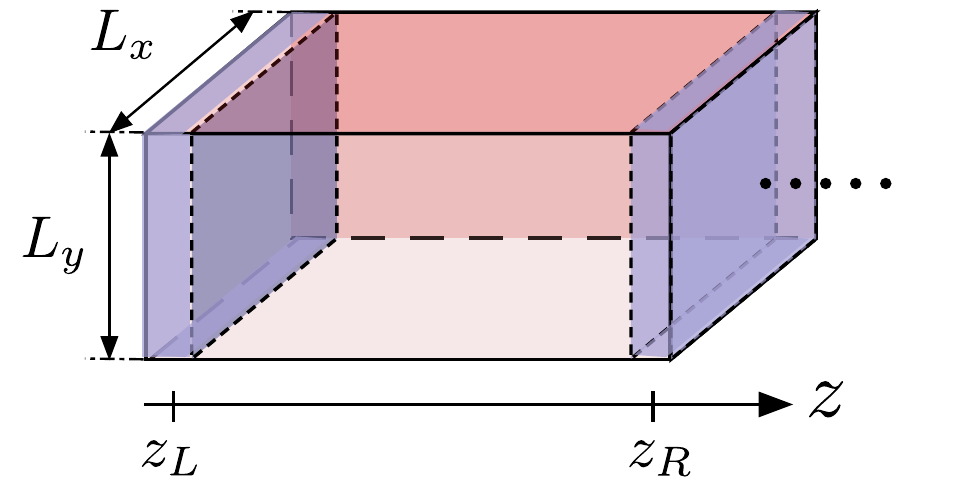}
\caption{Slab geometry of the system used in the recursive Green's function method. The system is bounded in the transverse directions. The system is extended rightward at each step of the simulation. In this process, the method calculates the matrix for the two-point Green's function involving points in the slab at $z = z_L$ and points on the rightmost region at $z = z_R$. The rightmost regions (indicated by a blue slab) may overlap with the updated region at the next computational step. In the Floquet setting, one of the transverse directions is the harmonic-space direction.}
\label{fig:figslab}
\end{figure}
\end{center}
%%%%%%%%%%%%%%%%%%%%%%%%%%%%%%
The recursive Green's function method was originally developed by Mackinnon and Kramer~\cite{mackinnon1983scaling} and adapted in Refs.~\onlinecite{aoki1985critical,huckestein1989new}. It was first employed to study Anderson localization in three dimensions numerically. In this method, one considers the system in a slab geometry; see Fig.~\ref{fig:figslab}. At every step $n$ of the computation, one calculates the inverse of two-point Green's function $A_n = \left[ \matrixel{z \in C_{z_0} }{G_{z_0; z_n} (\epsilon) }{z \in C_{z_n}} \right]^{-1}$, using the relation

\begin{align}
	A_{n+1}=(E-H_n)A_n-A_{n-1},
	\label{eq:eqrecur}
\end{align}

where $H_n$ is the Hamiltonian defined only on the $(d-1)$ dimensional slice (of the $d$-dimensional system) centered at $z = z_n$; this slice is denoted by $C_{z_n}$. $G_{z_0; z_n} (\epsilon)$ represents the resolvant matrix $\left( \epsilon - H_{z_0; z_n} \right)^{-1}$ for a system with open boundaries at $z = z_0$ and $z = z_n$, at energy $\epsilon$.  

The localization length $\xi (\epsilon)$ at energy $\epsilon$ can then be computed from the relation
\begin{align}
\xi^{-1} (\epsilon) =-\lim_{n\rightarrow \infty} \frac{\ln \text{Tr} \left[\abs{G_{z_{0}; z_{n}} (\epsilon + i 0^+)}^2 \right]}{2(z_n-z_0)}.
\label{eq:loclength}
\end{align}

%$G_{1n}$ is the sub block of the full Greens function or the resolvent matrix that couples the first and the $n$th (last) site and it can be obtained by iterating the following system of recursive equations,
%\begin{align}
%	G_{1,n+1}=G_{1,n}G_{n+1,n+1},\,\, G_{n+1,n+1}=\frac{1}{[E-H_{n+1}-G_{nn}]}\nn
%\end{align}
%he Greens function components determines $U$ by $U_{n+1}=G_{1n}^{-1}$, $U_1=\mathbb{I}$,  and $U_0=0$. 

%Numerical implementation of these recursion relations requires appropriate rescaling to extract the smallest eigenvalue of $A_n$ which rises exponentially in $n$ for a localized system. 

%The rescaling is performed to add numerical stability to the calculation of the localization length $\xi$, and prevent the entries in the matrix $A_n$ from growing too large.
In practice, for determining the localization length $\xi$, it is useful to perform a singular value decomposition of $G_{z_0; z_n}$ to extract the largest few eigenvalues (which correspond to the smallest eigenvalues of $A_n$), and rescale the values so as to avoid having entries that are all zero due to falling below machine precision. For a localized system, this is extremely important because the eigenvalues of $G_{z_0; z_n}$ fall off exponentially in the length of the system. 

%This is important since the smallest eigenvalue of $U_n$ is the most dominant contribution to $G_{1n}$ and by extension $\xi$. 

\subsection{Use of the recursive Green's function approach in the Floquet setting}

The above method is very naturally extended to the Floquet setting when we view the Hamiltonian $H^{ij}_m$ as an effective hopping Hamiltonian in $(d+1)$ dimensional space. Then, the slabs shown in Fig.~\ref{fig:figslab} are $d$-dimensional slices with one (transverse) dimension being the harmonic-space direction. Nevertheless, there are a few subtleties in making this extension which we now discuss. 

First, what is the physical meaning of the localization length $\xi_{d+1}$ computed using Eq.~(\ref{eq:loclength}) in this setting? For a given Floquet eigenstate $\ket{\psi_{\alpha, n} (t)}$, the probability of finding a particle in the location $i$ (corresponding to physical Hilbert space basis state $\ket{i}$) \emph{over the course of a full period}, is given by   

\begin{align}
\frac{1}{T} \int^T_0 \abs{\braket{i}{\psi_{\alpha, n} (t)}}^2 = \sum_m \abs{\phi^{(\alpha, n)}_{i,m}}^2 \equiv \abs{\phi^{(\alpha)}_{i} }^2
\end{align}

Note that given the redundancy of solutions, the final amplitudes $\abs{\phi^{(\alpha)}_i}^2$ are independent of the harmonic space index $n$ in $\ket{\psi_{(\alpha, n)} (t)}$. Now, these amplitudes can next be used to sensibly define a Floquet localization length $\xi_F$ through their exponential decay in space for a fixed eigenstate. 

The computation of Eq.~(\ref{eq:loclength}) extended to the Floquet setting must directly probe the decay of these amplitudes. To see that this is the case is straightforward. Expanding the Floquet Hamiltonian used in the definition of the resolvant matrix as $H^F_{z_0; z_n} = \sum_{n, \alpha} \left( \epsilon_\alpha + n \omega \right) \ket{\psi_{\alpha, n}} \bra{\psi_{\alpha,n}}$, it is easy to show

\begin{align}
\text{Tr} & \left[\abs{G_{z_{0}; z_{n}} (\epsilon + i 0^+)}^2 \right] \nonumber \\
&= \pi \sum_{\alpha, m, i, j} \abs{\phi^{(\alpha)}_{i\in C_{z_0}} }^2 \abs{\phi^{(\alpha)}_{j\in C_{z_n}} }^2 \; \delta (\epsilon - \epsilon_\alpha + m \omega). 
\end{align}

Thus, in the Floquet setting, the method produces the same two-point spectral function as would appear in the usual static setting, with wavefunction amplitudes replaced by $\abs{\phi^{(\alpha)}_i}^2$ as we desire. 

Another subtlety particular to the Floquet setting comes from the inherent redundancy of the Floquet eigenstates. Suppose $h^{i,j}_m = 0$ for $m \neq 0$, as for a static system. In this case, the Floquet eigenstates are simply replicas of the eigenstates of the static Hamiltonian, but at energies shifted by multiples of $\omega$. If we are then interested in computing the localization length of eigenstates of this static Hamiltonian at energy $\epsilon$, it will get spurious contributions from states at energy $\epsilon + n \omega$, which may have wildly different localization lengths (and may be even delocalized in the case where a mobility edge exists). This problem can be circumvented by making the following reasonable approximation: we limit the harmonic-space dimension to $N = c A / \omega$, where $c$ is an $\mathcal{O} \left( 1 \right)$ constant, and $A$ is the drive amplitude. This is justified because Floquet eigenstates are in fact confined to a width of about $\sim A/\omega$  due to the tilt of the Floquet lattice in the harmonic-space direction. With this approximation, we can obtain the Floquet localization length of states at the energies $\epsilon$ that span the \emph{original spectrum of the static Hamiltonian}, with the resolution $N \omega = c A$. 

We implemented this algorithm for the case of a periodically driven one dimensional Anderson insulator~\cite{agarwal2017localization}. In the present work, we extend this method to the case of driven disordered models with mobility edges. The two cases of interest are driven random dimer model (DRDM) and driven random Landau model (DRLM). %These two models possess sub-extensive number of delocalized states with one crucial difference. The extended states in DRLM is topologically protected while the extended states in  DRDM is not protected. 

 \section{Driving the random Landau model}
 
 We now use the methods above to ascertain the role of driving on the localization properties of electrons in the disordered lowest Landau level.  
 
 \subsection{Review of the static model}
 
We begin by revisiting the static RLM. The localization properties of the disordered lowest Landau Level (LLL) was first studied by Ando and Aoki~\cite{aoki1985critical} using the recursive Green's function method\cite{mackinnon1983scaling} and assuming a uniform concentration of randomly placed delta-function scatterers. Huckestein et al.~\cite{huckestein1989new, huckestein1990one} instead modeled the local potential disorder as a random variable with zero mean and Gaussian correlations in space of some fixed amplitude. This potential can be efficiently generated and appears to yield results for the localization length that scales better with system size~\cite{huckestein1989new} than Ando's choice of delta-potential scatterers, and also allows efficient extraction of scaling exponents. We will thus use the formulation of Huckestein et al. in this study.  

We now outline the details of the model. We consider a system with length $L_x$ and width $L_y$ with periodic boundary conditions in the $y$-direction and with $L_x\rightarrow \infty$. The single-electron Hamiltonian in the Landau gauge, in the presence of disorder is given by

\begin{align}
H=\sum_{mk}E_{mk}|m, k\rangle\langle m,k|+	\sum_{mkm'k'}V_{mm',kk'}(\vec r)|m, k\rangle\langle m',k'|,
\end{align}
where $V_{mm',kk'}(\vec r)=\langle m,k|V(\vec r)|m',k'\rangle$ and basis states $\ket{m,k}$ represent states with fixed momentum $k$ in the y-direction. In particular,  
\begin{align}
\psi_{mk}=	\langle r|m,k\rangle=\frac{1}{\sqrt{L_y l_c}}e^{iky}\chi_n\left(\frac{x-kl_c^2}{l_c}\right).
\end{align}
with $\chi_{m}(x)=(2^m m!\sqrt{\pi})^{-1/2}H_m(x)e^{-x^2/2}$ where $H_{m}(x)$ is the $m$th Hermite polynomial, $l_c = \left( \hbar / eB \right)^{1/2}$ is the magnetic length, and the diagonal part $E_{mk}=\hbar \omega_c (m+\frac{1}{2})$ is a constant for a fixed Landau level (here $m = 0$) and is thus neglected. Note further that in every patch of width $l_c$ in the $x$-direction, there are $L_y$ such basis states. 

The potential matrix elements $\langle k_1|V(r)|k_2\rangle$ are non-zero for $k_1 \neq k_2$ due to disorder. These elements can be generated to  satisfy desired statistical properties as noted above, as outlined in Ref.~\onlinecite{huckestein1990one}. 
%The continuum version of the matrix elements for such a disorder potential for the LLL is given as,
%\begin{align}
%	\langle k_1|V(r)&|k_2\rangle=\frac{V_0}{\sqrt{2\pi L_y}}e^{-\frac{(k_1-k_2)^2l_c^2\beta^2}{4}}\times\nn\\&\int d\xi u_0\left(\beta \xi+\frac{(k_1+k_2)l_c}{2},(k_1-k_2)l_c\right)e^{-\xi^2}.
%\end{align}
%Where the $u_0$ is the gaussian correlated random variable satisfying $\overline{u_0(r,k)u_o(r',k')}=\delta_{k,-k'}\delta(x-x')$. The numerical version of the above expression written on a discretized grid is given by,
For completeness, the matrix elements are given by the relation

\begin{align}
\langle k_1|V(r)|k_2\rangle =V_0\frac{ e^{-\frac{(k_1-k_2)^2l_c^2\beta^2 }{4}\sum_p  u_{2k_1+p,k_2-k_1}e^{-\frac{\pi^2 l_c^2}{L_y^2 \beta^2}p^2}
}}{\sqrt{(2\pi)^{1/2} l_c L_y \sum_pe^{-2\frac{\pi^2 l_c^2}{L_y^2 \beta^2}p^2}}}\label{eq:staticRLMmatrixel}
\end{align}

where the elements $u_{i,j}$ are drawn independently from a normal distribution, $V_0$ sets the amplitude of the disorder, and we set $\beta = 1$ for which the correlation length of the disorder is equal to the magnetic length; it is thus a smoother version of the delta-function disorder potential. 
%All lengths are measured in units of $\sqrt{2\pi}l_c$. For the uncorrelated potential, $V_0$ is defined as $\overline{V(r)V(r')}=V_0^2\delta(r-r')$. 

The basis states $\ket{k}$ may be viewed as basis states of a one-dimensional hopping matrix, and in this sense, the above model is effectively a 1D Anderson impurity model, albeit with very special diagonal and off-diagonal disorder. The lattice sites of this model are marked by their position $i \equiv \frac{kL_y}{2\pi}$. The diagonal matrix elements $\langle k|V(r)|k\rangle$ are the random onsite energies $\epsilon_i$. In the above language, the model may appear to be a `long range hopping model' since the off-diagonal disorder couples sites distance $l_c$ apart in real space, or alternatively, sites that are $L_y$-distant, a number which is scaled to infinity in the thermodynamic limit. However, the amplitude of this off-diagonal matrix elements is also reduced appropriately so that $l_c$ is the only relevant quantity in the thermodynamic limit. This suppression allows truncation of next to nearest neighbor intercell terms. 
%This analogy to the Anderson model, albeit in the degenerate $k$ space, allows the use of recursive Green's function approach to numerically determine the localization length.
 Within a self-consistent Born approximation, the width of the LLL due to the disorder is given by $\Gamma = 2 V_0/\sqrt{2\pi} l_c$; all energies will be presented in units of $\Gamma$ in what follows. 
% with coupling between all states within the range $|i-j|\sim L_y/2\pi l_c$. This long range form is a consequence of the projection operation onto the LLL. For a finite $L_y$, the RLM model has only localized states. However, the largest localization length is of the order of $L_y$ and, in the two-dimensional limit, $L_y\rightarrow \infty$, both the range of the hopping matrix elements and the maximum localization length diverge. In contrast to the usual Anderson models, the localization properties can only be extracted through system size scaling. 

\subsection{Formulation of the driven Random Landau Model}

% In the following, we introduce periodic driving to the model. 

%%%%%%%%%%%%%%%%%%%%%%%%%%%%%%%%%%%%%%%%%%%%%%%%%%%%%%%%%%%%%%%%%%%%
\begin{figure}[tb]
  \centering
\includegraphics[scale=0.35]{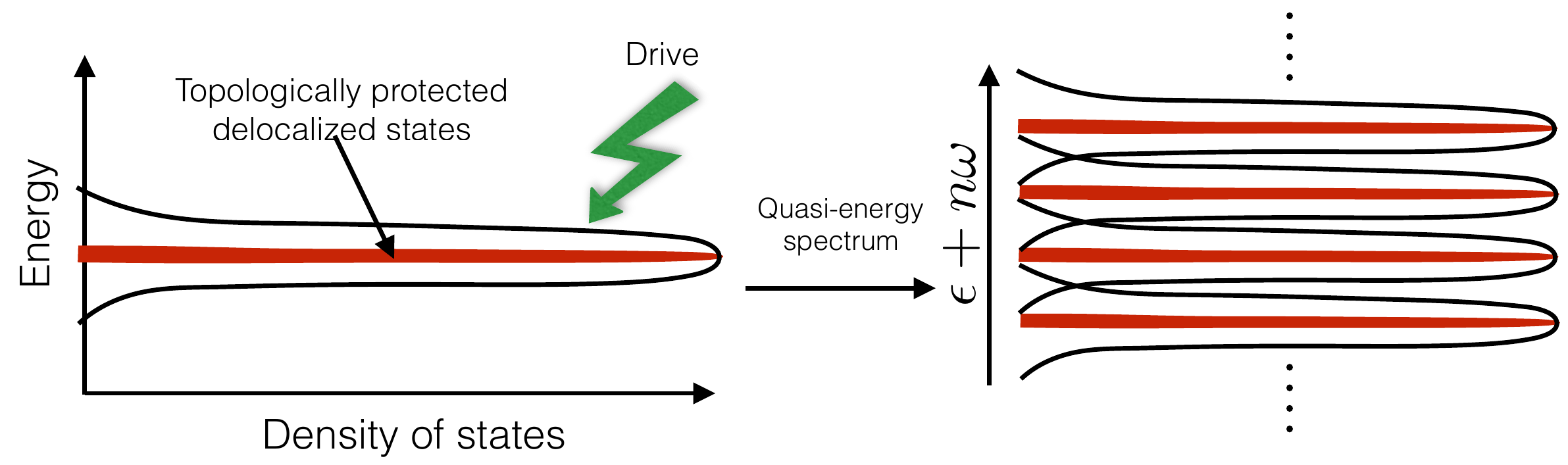}
\caption{Schematic of the lowest Landau level subject to a periodic drive. The resulting quasi-energy spectrum on the right creates the repeated pattern of the disordered Landau level band.}
     \label{fig:fllschematic}
\end{figure}
 %%%%%%%%%%%%%%%%%%%%%%%%%%%%%%%%%%%%%%%%%%%%%%%%%%%%%%%%%%%%%%%%%
 %%%%%%%%%%%%%%%%%%%%%%%%%%%%%%%%%%%%%%%%%%%%%%%%%%%%%%%%%%%%%%%%%%%%
\begin{figure}[tb]
  \centering
\includegraphics[scale=0.4]{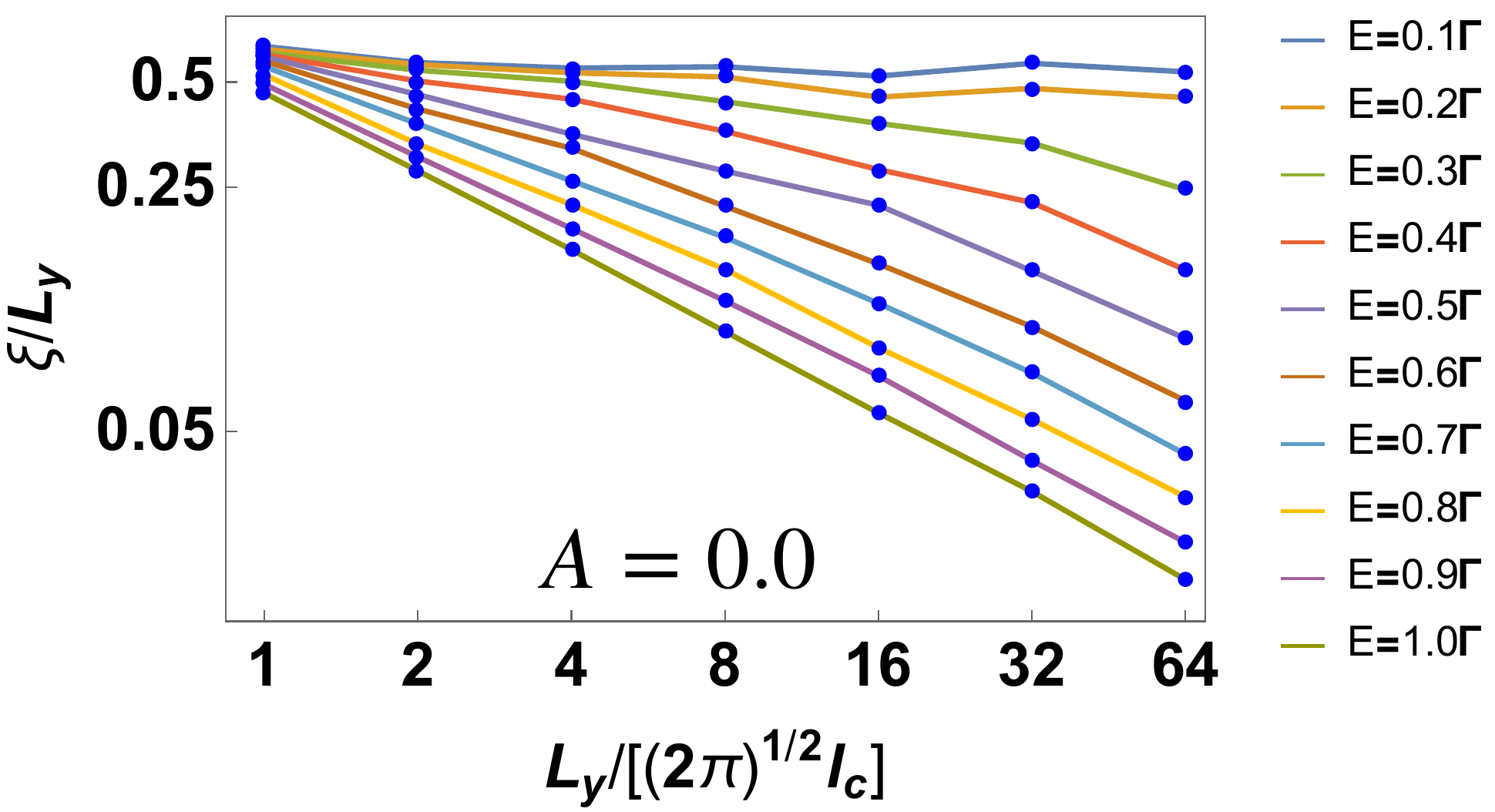}
\caption{Zero drive case ($A=0$): Dimensionless localization length $\xi/L_y$ plotted as a function of the dimensionless width $L_y/((2\pi)^{1/2}l_c)$ for different values of energy $E=0.1\Gamma-1.0\Gamma$. The number of harmonics is fixed at $n=6$. }
     \label{fig:rlmstatic}
\end{figure}
 %%%%%%%%%%%%%%%%%%%%%%%%%%%%%%%%%%%%%%%%%%%%%%%%%%%%%%%%%%%%%%%%%%%%
%%%%%%%%%%%%%%%%%%%%%%%%%%%%%%%%%%%%%%%%%%%%%%%%%%%%%%%%%%%%%%%%%%%%
\begin{figure*}[tb]
  \centering
\includegraphics[scale=0.5]{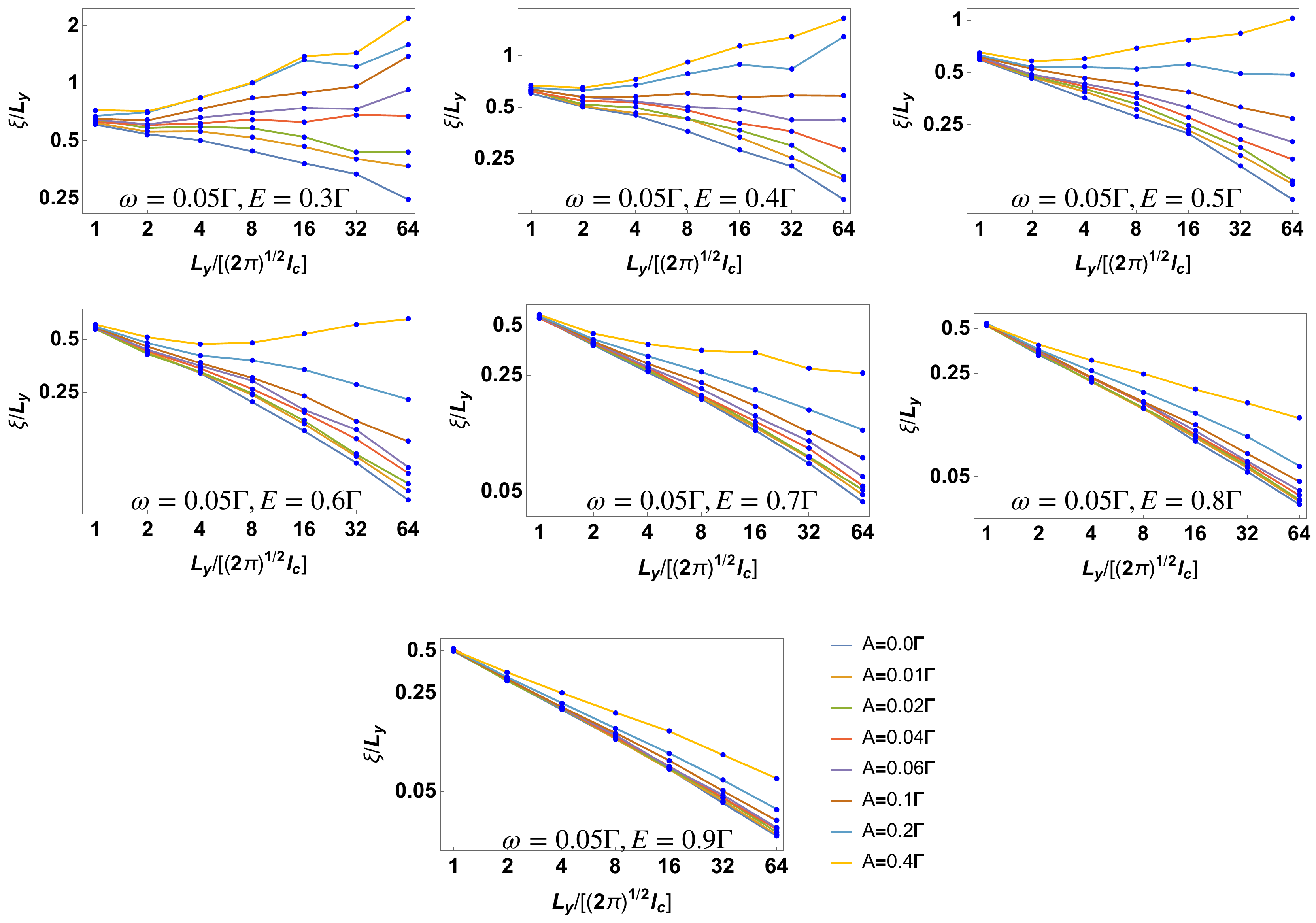}
\caption{Dimensionless floquet localization length $\xi/L_y$ plotted as a function of the dimensionless width $L_y/((2\pi)^{1/2}l_c)$ for different values of the drive amplitude $A=0 \Gamma, 0.01\Gamma, 0.02\Gamma, 0.04\Gamma, 0.06\Gamma, 0.1\Gamma, 0.2\Gamma, 0.4\Gamma$. The frequency of the drive  and the number of harmonics is fixed at $\omega=0.05\Gamma$ and $n=6$ respectively. The static energy values considered are$E=0.3\Gamma-0.9\Gamma$ in steps of $0.1\Gamma$. $E = 0$  and $E = \Gamma$ are the center and edge of the LLL band respectively.}
     \label{fig:fll}
\end{figure*}
 %%%%%%%%%%%%%%%%%%%%%%%%%%%%%%%%%%%%%%%%%%%%%%%%%%%%%%%%%%%%%%%%%%%%

Floquet dynamics of the driven RLM model (DRLM) can be studied by introducing a periodic driving term to the disordered Hamiltonian projected to the LLL. We consider a time dependent monochromatic drive of the form

 \begin{align}
V_D(r,t)=4E_0 \cos(\omega t+ k_x x)
\label{eq:driveRLM}
\end{align}

with period $T \equiv 2\pi/\omega$, and sinusoidal variation along the x-direction with wave-vector $k_x = \pi/l_c$. Note that spatial variation of the potential is necessary to obtain a finite response and the magnetic length is the natural length-scale for this purpose; we note that the precise value of $k_x$ is not qualitatively important. The drive amplitude $4E_0$ is assumed to be much smaller than the cyclotron gap (otherwise, LL transitions would have to be considered), but can be comparable to the disorder strength $V_0$. %Such a driving term can be thought of as a temporal generalization of a periodic lattice potential considered by Huckestein~\cite{huckestein1994corrections} $V_P(r,t)=4E_0 \cos(\sqrt{2} \pi x/a)\cos(\sqrt{2} \pi y/a)$. The special values of the period $a$ that split the Landau level into $q$ sub-bands can be written as $2\pi l_c^2/a^2=q/p$ or $\frac{a^2}{l_c^2}=2\pi \frac{p}{q}$. 

The projected matrix element for the driving potential is a diagonal term given by

\begin{align}
\matrixel{k_1}{V_D(r,t)}{k_2} = A\cos (\omega t+ k_1 k_xl_c^2) \delta_{k_1,k_2} 
\label{eq:driveRLMmatrixel}
\end{align}

 where the overall drive amplitude is given by $	A=4E_0 e^{-\frac{1}{4} l_c^2 k_x^2}$. The diagonal form is a consequence of the choice of the x-dependent phase modulation in the drive potential and can be generalized by considering a more general spatial variation that depends on both $x$ and $y$. This again does not change the result qualitatively and thus we restrict ourselves to the diagonal drive. The DRLM model is thus given by
 
\begin{align}
H=	\sum_{k_1k_2}\langle k_1|V(r)+V_D(r,t)|k_2\rangle|k_1\rangle \langle k_2|
\end{align}

with matrix elements given in Eqs.~\ref{eq:driveRLMmatrixel} and~\ref{eq:staticRLMmatrixel}. 

%Next, in the Floquet lattice representation, the Hamiltonian as a Floquet lattice model. The effective one dimensional nature with long range correlated disorder of the RLM model in the degenerate subspace can be enlarged to include the driving terms in the form of additional dimension. 

Finally, the Floquet lattice representation of the DRLM model is given by
\begin{align}
& H=\nn \\&\sum_{k,n}\frac{A}{2}\left(e^{i k_x k l_c^2}|k,n\rangle\langle k,n+1|+e^{-i k_x k l_c^2}|k,n+1\rangle\langle k,n|\right)\nn\\+&\sum_{k_1,k_2,n}(n\omega\delta_{k_1,k_2}+V_{k_1,k_2}(r))|k_1,n\rangle\langle k_2,n|, 
\end{align}

 where the index $n$ corresponds to the harmonic space dimension as usual. The Hamiltonian has $D$ unique eigenfunctions corresponding to the LLL degeneracy $D$. All other eigenstates are constructed by translation of this $k$ space lattice in harmonic space with a corresponding increase in energy by appropriate multiples of the drive frequency ($\omega$). We numerically compute the localization length associated with Floquet lattice model $H$ using recursive Green's function method as discussed in Sec.~\ref{sec:rgf}. 
  
  %developed by McKinnon and Kramer for nearest neighbor hopping term. Localization length in the static RLM was computed as an extension of the original McKinnon and Kramer method, by generalizing to Hamiltonians with long range hopping. In terms of the Green's function $G(n) = (E-H_F(n))^{-1}$ connecting the $N$ harmonic-space sites at one end of the k-space lattice to those at the other end, the Floquet localization length $\xi$ as a function of energy and width of the sample is given by $\xi^{-1}=-\lim_{n\rightarrow \infty} \frac{\ln Tr [G G^{\dagger}]}{2(n-1)}$.

\subsection{Numerical results}

Extracting the localization length in the DRLM model is computationally challenging due to the two-dimensional nature of the system. In particular, the localization length is first found for fixed $L_y$ and subsequently scaled to the thermodynamic limit. Increasing $L_y$ increases the linear dimension of the Green's function matrix between the `left' and `right' end of the system (see Fig.~\ref{fig:figslab}). In the one-dimensional picture of this model, increasing the Hilbert space with $L_y$ corresponds to having $L_y$ (in units of $\sqrt{2 \pi} l_c$) states inside a strip of width $l_c$. This makes the scaling fairly challenging for the static case (see Refs.~\onlinecite{aoki1985critical, huckestein1990one}) although exponents have been reliably extracted with relatively small $L_y$. In our case, the computational complexity is further increased due to the additional harmonic space dimension.

%We follow the same procedure as the case of DRDM model, but for DRLM we need to infer the localization properties from how the localization length of the Floquet lattice scales with the transverse width of the system. This approach is similar to the one employed to extract the localization properties of the three dimensional Anderson localized systems~\cite{mackinnon1983scaling, kramer93}  and scaling properties of quantum Hall plateau transition~\cite{aoki1985critical, huckestein1990one}. The RGF method applied to the momentum space RLM model is different from the real space lattice models. 

%Here we need to divide the strip of width $L_y$ into cells and neglect the next to nearest neighbor inter-cell coupling. The size of each cell size is equal to the number of states $M=L_y/(\sqrt{2\pi}l_c)$. For the DRLM model this cell size is enlarged to $n\times M$, where $n$ is the number of harmonics. 

As a preliminary analysis, we recover the localization length as a function of $L_y = 1, 2, 4, 8, 16, 32, 64$ (in units of $\sqrt{2\pi} l_c $) for the static case $A=0$, and use the same harmonic space size, $n=6$, as used in the DRLM results.  The localization length tends to a fixed value as $L_x$ is increased (due to self-averaging); all results are presented for $L_x = 5000 l_c$. Fig.~\ref{fig:rlmstatic} shows the localization length as a function of the strip width $L_y$ for different values of energy $E$ that ranges from relatively close to the center of the band $E=0.1\Gamma$ to the edge of the band $E=1.0\Gamma$. The localization length clearly decreases as a function of $L_y$ away from the center of the band indicating localization as $L_y\rightarrow \infty$. However, as we approach the band center ($E=0.1\Gamma$), the localization begins to scale linearly with the system size (i.e. roughly constant $\xi/L_y$), indicating the onset of delocalization(see Fig.~\ref{fig:rlmstatic}). These results agree well with those of Huckestein et al. in Ref.~\onlinecite{huckestein1989new, huckestein1990one}. 

We now proceed to study the effects of driving on localized states in the band. Similar to the static case, a decreasing dimensionless localization length $\xi/L_y$ with increasing width $L_y/((2\pi)^{1/2}l_c)$ implies localization and a localization length increasing with the width is an indicator of delocalization. We focus on the energy range $0.3\Gamma<E<0.9\Gamma$ where the non-driven ($A=0$) case shows clear localization (see Fig.~\ref{fig:rlmstatic})~\cite{aoki1985critical, huckestein1990one}. The main question of interest is if these localized states are delocalized due to periodic driving. We restrict the number of harmonics to be $n=6$. Having too large value of $n$ can spuriously indicate delocalization just by the replication of bands in the quasi-energy space. At the same time, for a fixed $n$, the largest drive amplitude $A$ we can study satisfies $A \lesssim n \omega$. The choice of drive amplitude and the frequency considered in the work is dictated by these restrictions. All results have an energy resolution of $\sim n \omega \approx 0.3 \Gamma$; this resolution further dictates the closest energy from the band center that we can examine reliably.  

The results presented in Fig.~\ref{fig:fll} demonstrate that for energies $E=(0.3, 0.4, 0.5, 0.6 )\Gamma$, the localization length appears to grow larger with increasing $L_y$ for large drive amplitudes, whereas for the smallest drive amplitudes, the reverse is true (just like the static model case as a function of the static energy). For $E>0.6\Gamma$ the localization length increases with the drive amplitude but the drive is not strong enough to mix the delocalized modes with the localized modes at this energy scale effectively. While our finite size results with limited number of harmonics indicate that modes near the center become delocalized with sufficient drive, certainty about whether this actually happens for a finite range of energies in the thermodynamic limit would require larger sizes than we are able to do. It would also be very interesting to study if the exponent determining the scaling of the localization length with energy changes as a result of the driving. Again, simulations on much larger systems would be required to answer this question conclusively.

%As we will see, This is in stark contrast with the case of DRDM where the driving decreased the localization length and even localized the delocalized states. We infer that the topologically protected states delocalize even the localized states upon driving. Due to the limited system size access and the higher dimensional nature of the problem, we are not able to extract the exact scaling behavior as a function of parameters $A, \omega, E$. 

\section{Driven Random Dimer Model}

\subsection{Review of the static random dimer model}

  The static random dimer model~\cite{dunlap1990absence} (RDM) is a model of binary disorder where pairs of consecutive lattice sites (or dimer pairs) have a different onsite potential compared to the remainder of the system. For concreteness, we take this different value to be $\epsilon_b$ while non-dimerized sites have onsite potential $\epsilon_a \equiv 0$. The locations of these dimers are picked at random. The RDM has a spectrum which features a sub-extensive set of extended modes near the energy $E = \epsilon_b$ as long as $-2t < \epsilon_b < 2t$ where $t$ is the hopping amplitude. The localization length diverges as $\xi(E) \sim 1/\abs{E - \epsilon_b}^\nu$ with a critical exponent $\nu \approx 2$, which is not too dissimilar to the exponent $\nu_{\text{QH}} \approx 2.5$~(see Refs.~\cite{slevin2009critical, zhu2019localization} and references therein for critical exponent for the quantum Hall plateau transition) for the integer Quantum Hall system. Thus, the static RDM has localization properties that are reminiscent of the disordered quantum Hall system but with one significant difference: critical energy state in the RDM is not topologically protected. 
  
  \subsection{Floquet dynamics of the driven model}

With this difference in mind, it is sensible to ask the following question: can localized states in the RDM couple to the extended states in the vicinity of the critical energy due to the driving, and thus become delocalized. Specifically we study the following Hamiltonian 

\begin{align}
H_{\text{DRDM}} &= - t \sum_i c^\dagger_i c_{i+1} + \text{h.c.} + \sum_{i \in X} \epsilon_b \left[  c^\dagger_i c_i + c^\dagger_{i+1} c_{i+1} \right] \nonumber \\
&+ A \cos (\omega t) \sum_i (-1)^i c^\dagger_i c_i
\end{align}

where $X$ is a random set of non-contiguous sites that denote the first site in the dimer pair, $A$ is the drive amplitude and $\omega$ is the drive frequency. In consistency with previous work, we use a driving potential that alternates in sign from site to site. We assume that any site has probability $q$ to appear in the set $X$ however we disallow dimers from overlapping or being contiguous to one another.  

In the large frequency regime, $\omega/t \gtrsim 1$, the driving has the effect of reducing the coherence of the hopping process, $t \rightarrow t \mathcal{J}_0 (2 A / \omega)$, as is known from performing a Peierls transformation, which transfers the time-dependence of the drive on to the fermionic operators---the reduction follows from the time-averaging of the time-dependence of the hopping over one time period (see for example Ref.~\onlinecite{bairey2017driving}). In this case, we expect a localization transition to occur by which the extended states get localized when the condition $-2 t \mathcal{J}_0 (2 A / \omega) < \epsilon_b < 2 t \mathcal{J}_0 (2 A / \omega)$ is violated. (In fact, due to the oscillatory nature of the Bessel function, one expects a series of localization-delocalization transitions.)

\begin{center}
\begin{figure}[]
\includegraphics[width=3.5in]{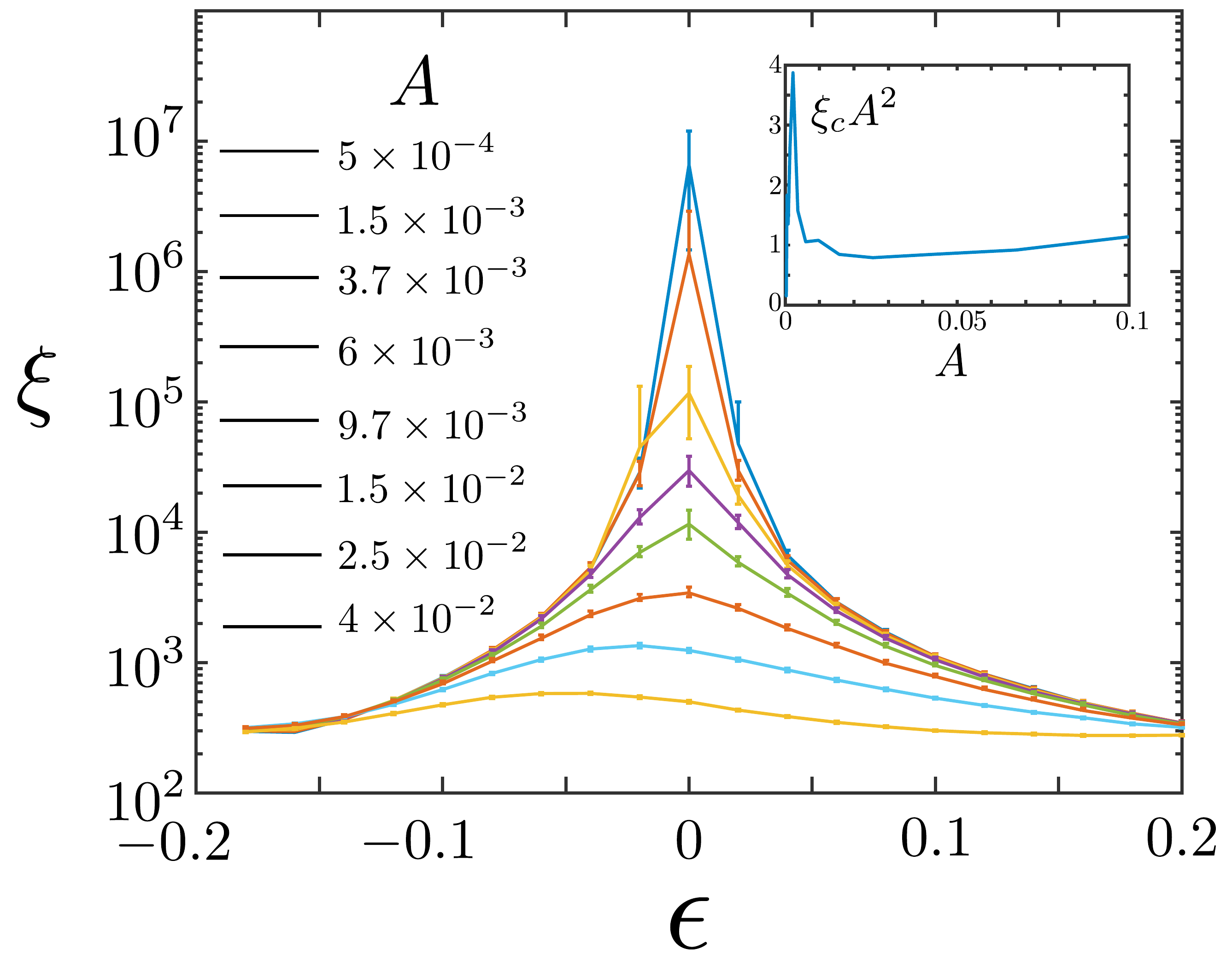}
\caption{Floquet localization length as a function of energy $\epsilon - \epsilon_b \in (-\omega/2, \omega/2)$ where the driving frequency $\omega = 0.4$. Note the logarithmic scale on the y-axis. All quantities are measured in units of $t \equiv 1$. The curves corresponding to smaller values of the localization length corresponding to progressively larger values of the drive amplitude $A$---an immediate collapse of the delocalization of the extended state can be observed as the amplitude is turned on. Inset: The localization length $\xi_c$ of the central mode which is extended in the absence of driving is seen to scale approximately as $\sim 1/A^2$ as per arguments in the main text; note that the deviation from the value $1$ at the smallest values of $A$ is well within the error bounds in calculating $\xi$ as can be seen from the main plot.}
\label{fig:DRDM2}
\end{figure}
\end{center}

Here again we focus on the low-frequency (and strong-driving) regime, where the frequency $\omega \ll t$. We can again map the dynamics of the system to that of a higher dimensional Floquet-Hamiltonian wherein hopping in the harmonic-space dimension is set by $A$. We use this Floquet representation to numerically evaluate the Floquet localization length as discussed above. The results are shown in Fig.~\ref{fig:DRDM2}. Most significantly, unlike the random Landau model where the delocalized mode is clearly stable to driving and the localization length of all modes is seen to increase due to driving, here we find that the localization length immediately begins to decrease as a drive is turned on. 
%%%%%%%%%%%%%%%%%%%%%%%%%%%%%%%%%%%%%%%%%%%%%%%%%%%%%%%%%%
%%%%%%%%%%%%%%%%%%%%%%%%%%%%%%%%%%%%%%%%%%%%%%%%%%%%%%%%%%
Due to the lack of topological protection in this instance, a conventional scattering analysis can explain our findings, at least at small amplitudes $A \ll \omega$. Let us focus on the `unscattered eigenstates' at momentum $k = \cos^{-1} (\epsilon_b/ 2t)$. A wave-packet centered at this momentum travels ballistically at a velocity $v \approx \frac{\partial \epsilon_k}{\partial k} \big|_{k = k_0}$, where $\epsilon_k = - 2 t \cos (k)$ is the dispersion of the clean band without the dimerized disorder. On time-scales $\tau \sim \text{min} \left[ 1/(vq), 1/\omega \right]$, we expect the wave-packet to essentially propagate ballistically. For $1/\omega \gg 1/(vq)$, scattering of the wave-packet will occur due to reflection off of imbalanced dimer pairs which have a local potential mismatch $\sim A$ between the dimer sites. (For no mismatch, the RDM would be recovered and there would be no reflection of such a wave-packet). On such time-scales, one can assume that the potential configuration is static, and solve the eigenvalue equation for a wave-packet propagating at energy $\epsilon_{k_0}$ and momentum $k_0$, in a flat potential landscape with a single imbalanced dimer. A straightforward calculation reveals a reflection amplitude $\abs{R}^2 = \left( \frac{A}{2t} \right)^2$ for the wave off of the dimer. Thus, the mean free path of such a wave-packet, $\lambda \sim \frac{1}{q \abs{R}^2} = \frac{4 t^2}{q A^2}$. Alternatively, if $1/\omega \ll 1/(vq)$, the particle's environment is not static and it encounters changes in the potential $\sim A$ on the time scale $\sim 1/\omega$; this will again be associated with an approximate reflection amplitude $\sim 1/\abs{R}^2$. In this case, the mean free path is given by $\lambda \sim \frac{v}{\omega \abs{R}^2} \sim \frac{v t^2}{\omega A^2}$. Finally, since the localization length is of the same order as the mean free path in one dimension, the above scaling forms are also valid for the Floquet localization length. We verify that the decrease in the Floquet localization length follows the scaling $\sim 1/A^2$ in the limit $\omega \ll v q$ in the numerics, see Fig.~\ref{fig:DRDM2}. 

%In the following section, we will contrast the findings for the DRDM with the case where the sub-extensive delocalized states are topologically protected. We consider the case of a disordered lowest Landau level that is subjected to a periodic driving. 

\section{Conclusions and Outlook}

In this paper, we investigated the effect of periodic driving on two disordered models that possess an isolated critical energy and associated subextensive but diverging number of quasi-extended modes in the thermodynamic limit. The two models are periodically driven  Random Landau model (DRLM) and Random Dimer Model (DRDM). The critical energy state in the former is topologically protected but not in the latter. We mapped the periodically driven systems to Floquet lattice models in one higher dimension, where the extra dimension corresponds to harmonic space. We then numerically studied the localization properties of this higher dimensional lattice using the recursive Green's function method. Our results indicate that the localization properties upon driving crucially depend on the origin of the the critical energy giving rise to delocalized behavior. For the DRLM case, the scaling of localization length as a function of sample width indicates an increase of localization length and an enhanced spectral range of delocalized modes as a function of drive amplitude.  As mentioned above, this may be understood as a consequence of the fact that the Chern number of the isolated band studied cannot be changed by the driving, which guarantees the presence of an extended state~\cite{laughlingquantized}. On the other hand, even weak driving of DRDM results in localization of the  delocalized mode. The localization of the delocalized mode is in contrast to the perturbative intuition that the mixing of localized and delocalized states must result in delocalization. These two contrasting behaviors points out that the fate of the mixing of localized and delocalized states upon driving can depend starkly on the origin of the delocalized states.

This contrasting behavior is similar to the contrasting behavior found\cite{hyman1996random, yang1996effects} for the effect of disorder on two spin-chain models – the dimerized spin$-1/2$ chain which is in a gapped, topological Haldane phase, and the Majumdar-Ghosh model spin-$1/2$ chain with nearest neighbor and next nearest neighbor couplings, which has a broken symmetry ground state with a gap to excited states like the dimerized chain. While the gap of the former is maintained at small disorder, showing the robustness of the topological protection, the gap of the Majumdar-Ghosh model is destroyed by any amount of disorder. More generally, the fragility of the extended states in the random dimer model to localization are reminiscent of the fragility of symmetry-protected topological insulators to symmetry breaking perturbations.  

The full scaling analysis of the DRLM case as a function of system size, amplitude, frequency, and energy requires access to larger system sizes and we leave this effort for future work. In this paper, we restricted our analysis to the case where both the single-band model possess an isolated critical energy and associated subextensive but diverging number of quasi-extended modes in the thermodynamic limit. An analogous situation is encountered in the 1D tight-binding chain of non-interacting electrons with disordered nearest neighbor hopping~\cite{eggarter1978singular, soukoulis1981off}. This model is equivalent to the Dyson model~\cite{dyson1953dynamics, de2016generalized} and is known to have a Lyapunov exponent (inverse localization length) that goes to zero at the center of the band.

It would be interesting to consider the influence of periodic driving on mobility edges with an extensive number of delocalized states which arise in three-dimensional Anderson localization and certain quasiperiodic potentials in one dimension~\cite{ganeshan2015nearest}. Another direction would be to systematically understand the role of dimensionality of the static model and the presence of multiple incommensurate driving frequencies. The additional incommensurate drive frequencies can be treated by adding more harmonic space dimensions on the Floquet lattice. 

\section{Acknowledgements}

SG acknowledges support from NSF OMA-1936351 at CCNY. KA acknowledges support of the UK Foundation Grant at Princeton University and NSERC Grants RGPIN-2019-06465 and DGECR-2019-00011 during final stages at McGill University. KA and RNB acknowledge support from DOE-BES Grant DE-SC0002140 at Princeton University. SG and RNB also acknowledge the hospitality of the Aspen Center for Physics (ACP) where part of this work was carried out; ACP is supported in part by NSF grant PHY-1607611.

%%%%%%%%%%%%%%%%%%%%%%%%
%%%%%%%%%%%%%%%%%%%%%%%%

%%%%%%%%%%%%%%%%%%%%%%%%
%%%%%%%%%%%%%%%%%%%%%%%%
\bibliographystyle{my-refs.bst}
\bibliography{references.bib}

\end{document}